\documentclass[letterpaper,10pt]{article}

\RequirePackage{fix-cm}

\newcommand{\abs}[1]{\left\lvert#1\right\rvert}

\usepackage{graphicx}

\usepackage{url}

\usepackage{bm}
\usepackage{amsmath}

\usepackage[round]{natbib}

\usepackage{fullpage}

\makeatletter
\let\@fnsymbol\@arabic
\makeatother

\begin{document}
%
%
\title{Observations on the flow structures and transport in a warm-core ring in the Gulf of Mexico}

\author{Doug Lipinski\thanks{
Institute for Networked Autonomous Systems,\newline
Dept. of Aerospace and Mechanical Engineering,\newline
University of Florida, Gainesville, FL, 32611
} \and Kamran Mohseni$^{1,}$\thanks{
Dept. of Electrical and Computer Engineering,\newline
University of Florida, Gainesville, FL, 32611 \newline
mohseni@ufl.edu~~Tel: (352) 273-1834~~Fax: (352) 392-7303 
}}

\maketitle

\begin{abstract}
This study presents several new observations from the study of a warm-core ring (WCR) in the Gulf of Mexico based on the ECCO2 global ocean simulation. Using Lagrangian coherent structures (LCS) techniques to investigate this flow reveals a pattern of transversely intersecting LCS in the mixed layer of the WCR which experiences consistent stretching behavior over a large region of space and time. A detailed analysis of this flow region leads to an analytical model velocity field which captures the essential elements that generate the transversely intersecting LCS. The model parameters are determined from the WCR and the resulting LCS show excellent agreement with those observed in the WCR. The three-dimensional transport behavior which creates these structures relies on the small radial outflow which is present in the mixed layer and is not seen below the pycnocline, leading to a sharp change in the character of the LCS at the bottom of the mixed layer. The flow behavior revealed by the LCS limits fluid exchange between the WCR and the surrounding ocean, contributing to the long life of WCRs. Further study of these structures and their associated transport behavior may lead to further insights into the development and persistence of such geophysical vortices as well as their transport behavior.

\end{abstract}

\section{Introduction}

Despite the importance and prevalence of warm-core rings (WCRs) in the Gulf of Mexico (GoM), many of the details of the transport and small-scale coherent structures in these flows remain poorly understood. WCRs have a significant impact on oceanic transport and energy balances~\citep{Elliott:79a,Lewis:89a,Oey:08a} and can also influence weather patterns including hurricanes due to their interaction with the atmosphere. It is known that hurricanes may rapidly intensify when passing over a WCR~\citep{Hong:00a,Shay:00a,Scharroo:05a}. Additionally, the mixing behavior in a WCR differs significantly from that of the surrounding ocean and may greatly influence biological activity such as plankton blooms~\citep{Franks:86a,Biggs:92a}.

The primary purpose of this paper is to present some newly observed structures and their associated transport behavior in a warm-core ring in the Gulf of Mexico. We have chosen to focus on newly observed small-scale coherent structures seen in the near surface region (that is, the ocean-atmosphere boundary) of the WCR. To study the structures in these flows, we use the technique of Lagrangian coherent structures (LCS). This technique is used to identify structures relevant to the Lagrangian transport of fluid. These structures represent barriers to transport and often reveal flow behavior that may be difficult or impossible to detect with Eulerian measures.

Computing the three-dimensional LCS present in numerical simulations of a WCR has revealed a previously unobserved type of structure in the mixed layer. Due to the transversely intersection LCS structures that appear in this region, we will refer to these structures as ``checkerboard LCS''. The flow in the checkerboard LCS region is characterized by consistent stretching behavior over a long time period and a large spacial domain. A parcel of fluid placed in this region is found to elongate in the azimuthal direction and compress in the radial direction.

To better understand the mechanisms that generate the checkerboard LCS structures, we present a simple flow model which produces similar structures. This flow model consists of both hyperbolic and shear stretching components of the flow as well as a spatially periodic perturbation. The velocity field generates large, nearly uniform stretching in the flow. The perturbation then creates localized regions of increased stretching, generating the checkerboard LCS.

\section{Lagrangian coherent structures}
\label{sec:LCS}

The term Lagrangian coherent structures (LCS) has come to refer to a class of techniques used to identify coherent structures in aperiodic, finite time flows. In steady state or periodic systems, classical dynamical systems techniques can be used to identify hyperbolic fixed points and their stable and unstable manifolds as well as other invariant manifolds in the system. These structures may then be used to study the flow topology and the corresponding mixing and transport in the system. However, in systems with general time dependence the same techniques no longer apply. To address this shortcoming and identify coherent structures in systems with general time dependence, various methods have been proposed. One of the most popular and successful methods has been the use of the finite time Lyapunov exponent (FTLE) to identify regions of locally maximum stretching in the flow~\citep{HallerG:00a,Marsden:05g}. Intuitively, one expects that regions with qualitatively different dynamics will be separated by a thin region of very large Lagrangian stretching~\citep{HallerG:00a}. Fluid parcels that straddle the boundary between two regions will be greatly deformed and stretched over time as they separate into different regions.

The notion of defining coherent structures by the stretching at their boundaries was formalized in \citet{Marsden:05g} by defining LCS as ridges in the FTLE field. This method of defining and computing LCS has since proven to be very effective in many cases, but it is worth noting that there are alternative definitions as well. In particular, Haller and others have developed a variational formulation for computing LCS~\citep{HallerG:11a,HallerG:11b}, there are methods for finding ``distinguished'' trajectories~\citep{WigginsS:02a,MadridJAJ:09a}, and there there are methods for finding maximally invariant sets~\citep{Marsden:05b,PadbergK:09a}. In this paper, we will use LCS as defined in \citet{Marsden:05g}: ridges of the FTLE field.

LCS techniques are primarily useful for determining and examining the Lagrangian transport of a system. In the past, these techniques have been used to investigate unsteady separation~\citep{HallerG:04a} and the flow over an airfoil~\citep{Mohseni:08f}, transport in jellyfish swimming and feeding~\citep{Mohseni:09d,DabiriJO:09a}, three-dimensional turbulence~\citep{Green:07a}, atmospheric transport~\citep{LeKienF:10a}, and many other applications. LCS provide a way to precisely identify the extent or boundaries of coherent structures and, correspondingly, a way to quantify their impact on transport phenomena. Crucially, LCS identify \emph{barriers to transport} and therefore reveal the structure underlying mixing and transport in a flow.

LCS have been applied to many ocean flows with good results. LCS have been used to study optimal pollution mitigation~\citep{Marsden:07e}, transport in a wind-driven double gyre~\citep{Coulliette:01a}, identify mesoscale eddies~\citep{Beron-Vera:08a}, and even investigate the transport of oil from the Deepwater Horizon oil spill~\citep{Mezic:10a,Huntley:11a,Olascoaga:12a}. Most ocean studies using LCS have focused solely on surface flows and used two-dimensional LCS computations. The expectation has been that since oceanic flows are highly stratified and vertical velocities are typically orders of magnitude smaller than the horizontal velocities, two-dimensional computations are appropriate. However, the length scales in the vertical direction are also orders of magnitude smaller than those in the horizontal, leading to flow gradients in the vertical direction that may exceed those in the horizontal. Recently, researchers have begun to focus more directly on the impacts of three-dimensionality on ocean LCS, noting that three-dimensional effects may be critically important even if the vertical velocity component is small~\citep{Sulman:13a}.

To compute the LCS, it is necessary to first compute the FTLE field. The most common method for doing so involves seeding a region of the flow with a grid of passive drifter particles at some initial time $t_0$ and advecting these particles with the flow field for some integration time $T$. This gives an approximation to the flow map
\begin{equation}
\vec{\Phi}_{t_0}^T(\vec{x}_0) = \vec{x}_0 + \int_{t_0}^{t+0+T} \vec{v}(\vec{x}(t),t) dt.
\end{equation}
Once the flow map has been computed, the Cauchy-Green deformation tensor is computed as
\begin{equation}
\Delta = \left( \dfrac{d\vec{\Phi}}{d\vec{x}_0} \right)^* \left( \dfrac{d\vec{\Phi}}{d\vec{x}_0} \right),
\end{equation}
where $^*$ denotes the transpose operator. This tensor contains information about the geometric deformation of the flow. The FTLE is then given by
\begin{equation}
\sigma_{t_0}^T(\vec{x}_0) = \dfrac{1}{\abs{T}} \ln \sqrt{\lambda_{\max}}
\end{equation}
where $\lambda_{\max}$ is the maximum eigenvalue of $\Delta$. Note that the integration time $T$ may be either positive or negative so for any flow there are always two sets of LCS. For $T>0$ the LCS are typically repelling structures and for $T<0$ they are attracting. Additionally, the magnitude of $T$ should be chosen so that sufficient detail is resolved in the LCS. As $T$ is increased, more LCS are revealed, but if $T$ is too large the complexity of the resulting structures may be difficult to interpret.

Once the FTLE field has been computed, the LCS are often visualized by directly plotting the FTLE field. The LCS are defined as ridges in the FTLE field which are visible just as ridges are visible on a topographical map. Various mathematical definitions of ridges are available but this choice does not seem to greatly affect the resulting LCS. For well defined ridges, the LCS typically permit very low or negligible flux and can be thought of as denoting barriers to fluid transport~\citep{Marsden:05g}. If desired, one may explicitly extract the ridges from the FTLE field in an additional step to obtain the LCS.

In this study, we have used an efficient ridge tracing algorithm to directly compute the LCS ridge surfaces in the WCR. This method greatly speeds the LCS computations by detecting some initial points on the LCS ridges and then tracing the ridges through space. By avoiding computations away from the FTLE ridges, the algorithm reduces the computational complexity from $\mathcal{O}(\delta x^{-3})$ to $\mathcal{O}(\delta x^{-2})$ in three-dimensional domains~\citep{Lipinski:12a}.

\section{Observations}
\label{sec:obs}

The observations presented here are based on data from the ECCO2 global ocean simulation which is publicly available at \url{http://ecco2.jpl.nasa.gov/}. We focus on the structure and transport of a warm-core ring in the Gulf of Mexico as found in the ECCO2 simulation on 1 February 2010. WCRs periodically form in the GoM when the loop current in the eastern Gom pinches off in a closed ring which contains warm Caribbean water. These rings typically persist for about 7-13 months~\citep{Sturges:00} as they drift slowly across the GoM before dissipating in the western GoM~\citep{Hurlburt:80}.

In WCRs, the flow field is largely two-dimensional, with horizontal flow speeds on the order of 1 m s$^{-1}$ and vertical speeds three to four orders of magnitude smaller. However, the horizontal length scales are on the order of 100 km while the vertical lengths scales are a few hundred meters. In combination, this means that flow gradients in the vertical direction may be of the same order or larger than those in the horizontal direction. Because of this, even a small vertical motion may have a large impact on the trajectories of fluid particles and the full three dimensional flow structure must be considered when analyzing fluid transport in WCRs.

\begin{figure}
\centering
\includegraphics[width=.8\textwidth]{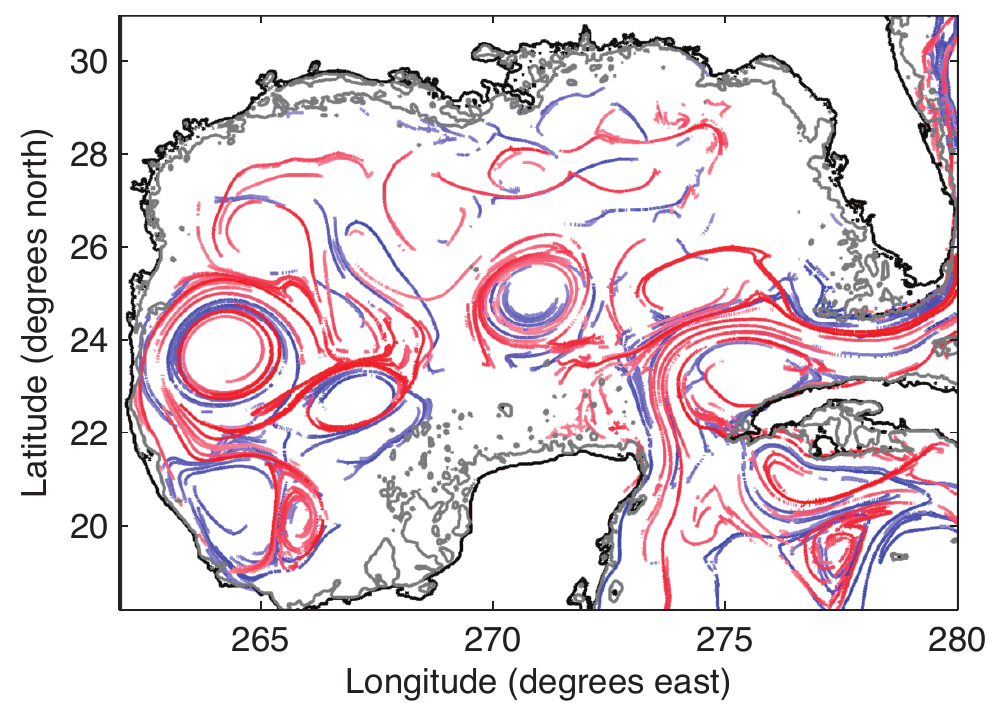}
\caption{A two-dimensional slice of the near-surface LCS in the Gulf of Mexico as computed from the ECCO2 dataset for 1 February 2010. Attracting LCS are shown in red and repelling are shown in blue. A large WCR is present in the center of the GoM and an older ring is in the western GoM. The loop current is also clearly visible.\label{fig:surfaceLCS}}
\end{figure}

A view of the near surface LCS in the GoM on 1 February 2010 is shown in figure~\ref{fig:surfaceLCS}. We have computed the three-dimensional LCS in the GoM using a ridge tracking algorithm to speed computations~\citep{Lipinski:12a}. An integration time of $T=\pm4$ weeks was used. This time was chosen to reveal the major structures in the flow. In figure~\ref{fig:surfaceLCS}, the loop current is clearly visible in the eastern GoM as it enters through the Yucatan Channel and exits through the Florida Straits. A recently shed WCR is in the central GoM and an older WCR is visible in the western GoM. Below, we will focus on the single WCR in the central GoM.

\begin{figure}
\centering
\includegraphics[width=1\textwidth]{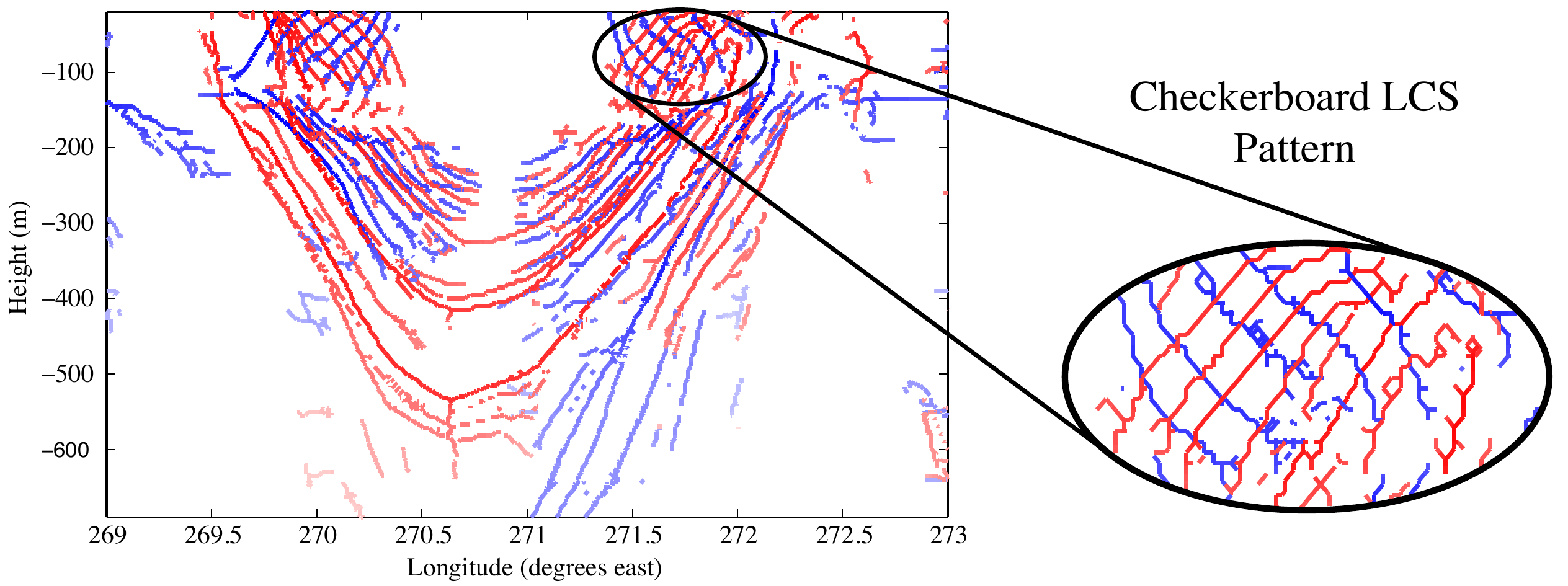}
\caption{A vertical cross section of the LCS in a warm-core ring in the Gulf of Mexico computed from the publicly available ECCO2 dataset. The inset shows the characteristic ``checkerboard'' pattern formed by the interaction of the attracting (red) and repelling (blue) LCS. \label{fig:WCRcheckerboard}}
\end{figure}

Figure~\ref{fig:WCRcheckerboard} shows a vertical cross section of the LCS in this WCR. There are several  features of note in this figure. First, the LCS reveal a closed bottom to the WCR. In this part of the WCR, the attracting and repelling LCS are nearly parallel and prevent transport in or out of the eddy. This limited transport is one of the reasons WCRs persist for so long. Additionally, this closed bottom indicates that the WCR has limited influence below a certain depth (about 550m in this case) and fluid below this depth does not get entrained or carried with the WCR. The closed bottom and finite depth of the WCR potentially allows for volumetric computations to determine precisely how much water is carried with the WCR and quantify the corresponding influence on heat, salt, and mass balances in the GoM.

Next, we note a clear difference between the LCS below about 120 m and those above with a sharp transition between these two regions. This 120 m depth marks the bottom of the mixed layer in the WCR. Below this depth, the attracting (red) and repelling (blue) LCS are approximately aligned with one another. However, in the mixed layer, the attracting and repelling LCS intersect transversely, forming a cross-hatched or ``checkerboard'' pattern. This checkerboard pattern is a new flow structure which has not been previously reported in geophysical flows.

\begin{figure}
\centering
\includegraphics[width=.32\textwidth]{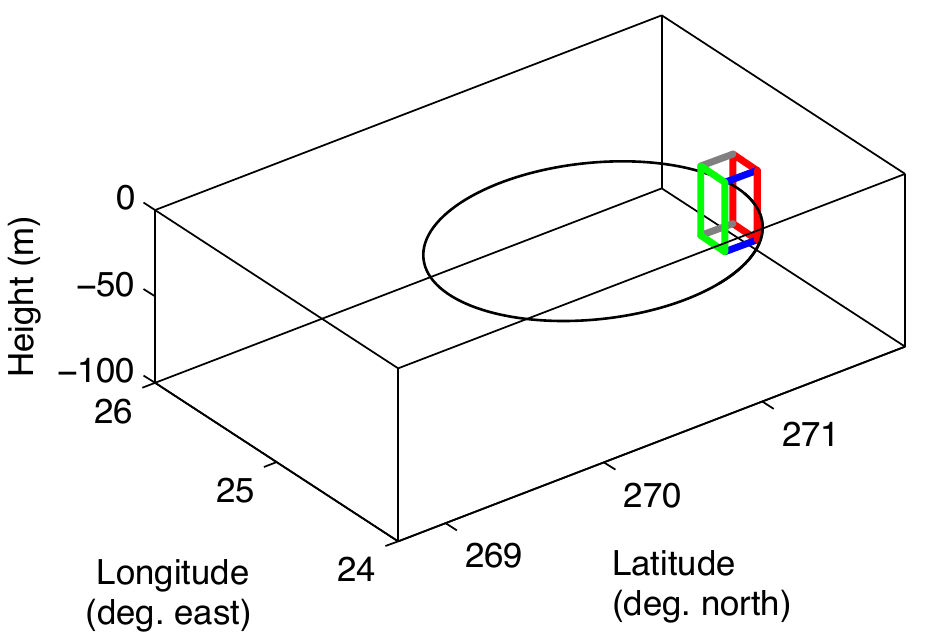}
\includegraphics[width=.32\textwidth]{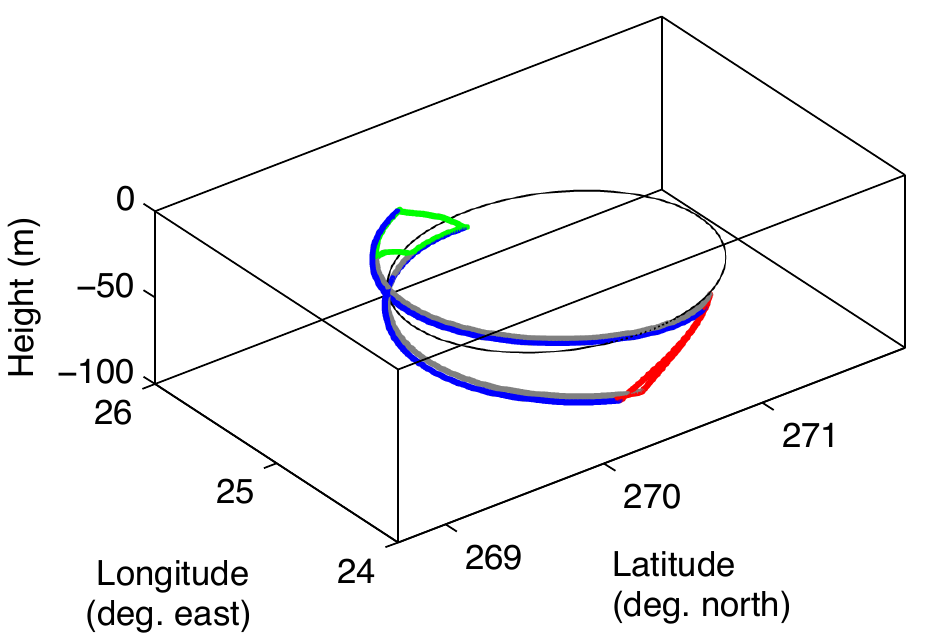}
\includegraphics[width=.32\textwidth]{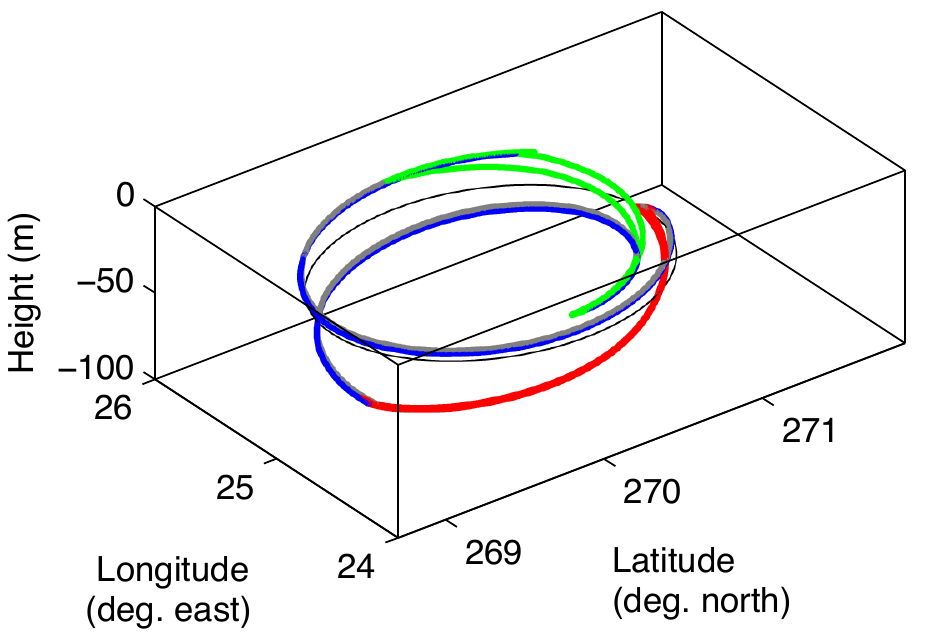}
\caption{A box of drifters placed in the checkerboard LCS region of a WCR. The drifter positions are shown at times of 0, 12.5, and 25 days. The black circle shows the approximate position of the center of the checkerboard region surrounding the warm-core ring.}
\label{fig:drifterbox}
\end{figure}

To investigate the flow behavior within the checkerboard region we place a box of passive drifter particles in this region of the WCR and track their motion over time. The particle positions are shown after 0, 12.5, and 25 days in figure~\ref{fig:drifterbox}. The box is initially $\approx$21 km across in the latitudinal and longitudinal directions and covers a depth range of 50 m. As can be clearly seen in figure~\ref{fig:drifterbox}, the initial box of drifter particles is quickly stretched and wrapped around the circumference of the WCR while being compressed in the radial direction. The box very quickly becomes greatly deformed by the horizontal velocities in the flow, but very little motion occurs in the vertical direction. Although it is difficult to see in figure~\ref{fig:WCRcheckerboard}, the box is also slowly compressed in the vertical direction and pushed upwards while being sheared in the radial direction with particles near the surface moving radially outward with respect to those below. The compression and upward motion is a small effect compared to the other motions observed. The near-surface radial outflow is expected due to friction-based disruption of the cyclo-geostrophic balance that is present in much of the WCR.

\begin{figure}
\centering
\hfill
\includegraphics[width=.48\textwidth]{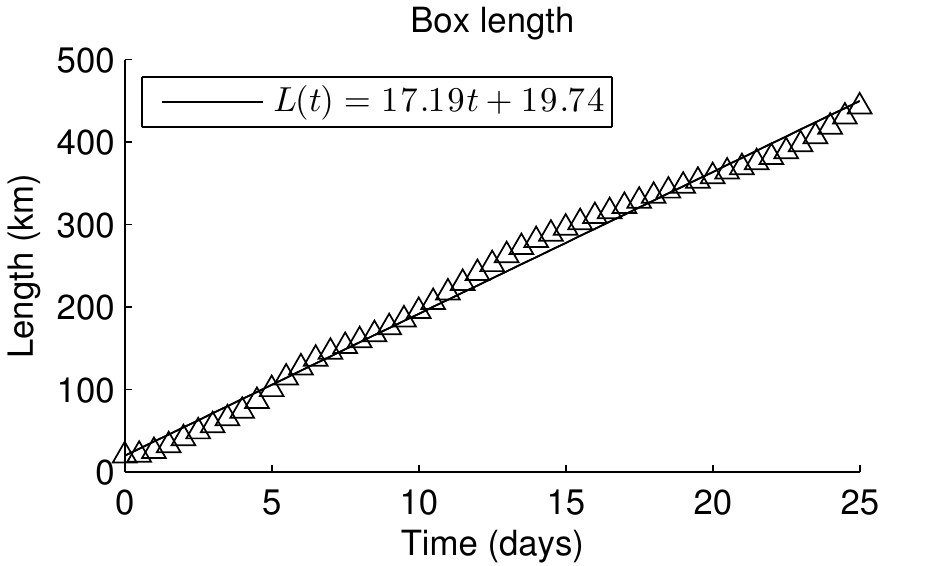}\hfill
\includegraphics[width=.48\textwidth]{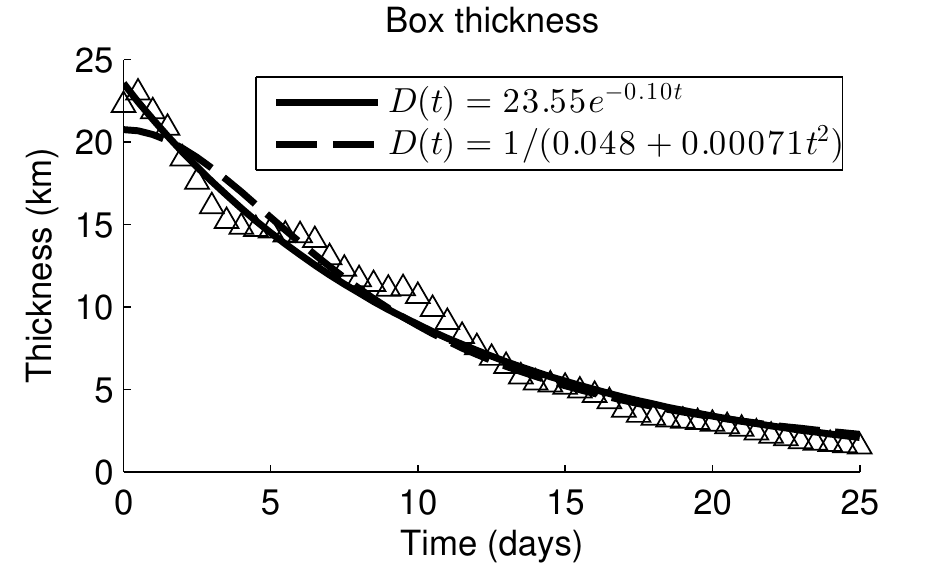}\hfill~
\caption{The deformed circumferencial length and radial thickness of the the drifter box shown in  figure~\ref{fig:drifterbox}. The length grows linearly at a rate of about 17.2 km day$^{-1}$ while the thickness decays approximately exponentially. As shown, a fit of $D(t)=1/(c_1+c_2t^{c_3})$ was also tried for the thickness, but the exponential provided a better fit. A least squares best fit was used to determine the regression curves.}
\label{fig:box_size}
\end{figure}

By carefully tracking the drifter particles it is possible to directly estimate the deformation caused by the flow. Since the box is quickly deformed from its initial configuration care must be used in such computations. Here, we focus on the circumferential length of the box and the radial thickness. The circumferential length is computed as the mean length of the elongating edges of the box, computed by integrating along each of these edges. The radial thickness is computed as the average of the distance between points on the longest two edges of the upper and lower faces of the box. To ensure accuracy, only the middle half of these edges is used. The results of this process are plotted in figure~\ref{fig:box_size}. The length of the box increases approximately linearly for the entire 25 day time period at a rate of approximately 17.2 km day$^{-1}$. The radial thickness decreases approximately exponentially over this same period, reaching a thickness of about 1.6 km after 25 days.

The flow deformation revealed in this analysis gives two additional insights. First, the observed stretching occurs over a relatively large region of space (an initially 21 km $\times$ 21 km $\times$ 50 m box) and time (25 days). The passive drifters that begin in this checkerboard LCS region remain in the checkerboard LCS region for the entire 25 day time period investigated and experience consistent deformation in both the circumferential length and radial thickness directions as shown in figure~\ref{fig:box_size}. Secondly, the velocity field appears to consist of three main components in the checkerboard region: (1) a large and sheared velocity component in the azimuthal direction (2) A small radial outflow near the surface (3) a small compression and upward motion in the vertical direction which must be balanced by expansion in the radial direction for conservation of volume.

\section{Checkerboard LCS model}
\label{sec:model}

As discussed in the previous section, the checkerboard LCS pattern is a new and prominent feature observed in this WCR. This pattern only appears in the mixed layer and there is a sharp transition at the bottom of the mixed layer from the transversely intersecting LCS above to the parallel LCS below. We attribute this sudden change to the greatly reduced radial and vertical flow components below the mixed layer. In the mixed layer, boundary interactions and wind forcing cause unique flow characteristics which can generate the checkerboard LCS pattern. Specifically, wind forcing influences the near surface velocity field through frictional affects and disrupts the cyclo-geostrophic balance of the WCR. This leads to a pressure gradient driven radial flow component due to the elevated sea surface height of the  WCR. This radial outflow component decreases quickly with depth, leading to relatively large values of shear rate in the radial flow. Combined with the shear in the much larger azimuthal velocity component, these two shear components in different planes acts to create stretching that depends on the direction of time integration. When a small perturbation is added, this generates the checkerboard LCS pattern seen in the WCR. A small amount of compression in the vertical direction was also observed, but it is omitted from the model since the resulting stretching was much smaller than that due to shear effects.

The model we use is a simple, non-time-dependent velocity field given by
\begin{align}
\dot{x} & = \dot{\gamma}_1 y, \notag \\
\dot{y} & = 0, \label{eq:model}\\
\dot{z} & = -\dot{\gamma}_2 x\left[1+\varepsilon\sin\left(\dfrac{2\pi}{\lambda}z\right)\right], \notag
\end{align}
where $\dot{\gamma}_1$ and $\dot{\gamma}_2$ are shear rates, $\varepsilon$ is a small perturbation magnitude, and $\lambda$ is the perturbation wavelength. Although this flow is given in Cartesian coordinates, $x$, $y$, and $z$ are analogous to the radial, vertical, and azimuthal directions in the WCR. The $z$ dependence is periodic just as the WCR is periodic in the azimuthal direction. The periodic perturbation is associated with deviations from axisymmetry in the WCR. To more closely match the velocity magnitudes seen in the WCR, constants could be added to the velocity components in equation~\ref{eq:model}. However, constant terms have no affect on the resulting FTLE values or LCS so they omitted here for simplicity.

\begin{figure}
  \centering
  \includegraphics[width=0.65\textwidth]{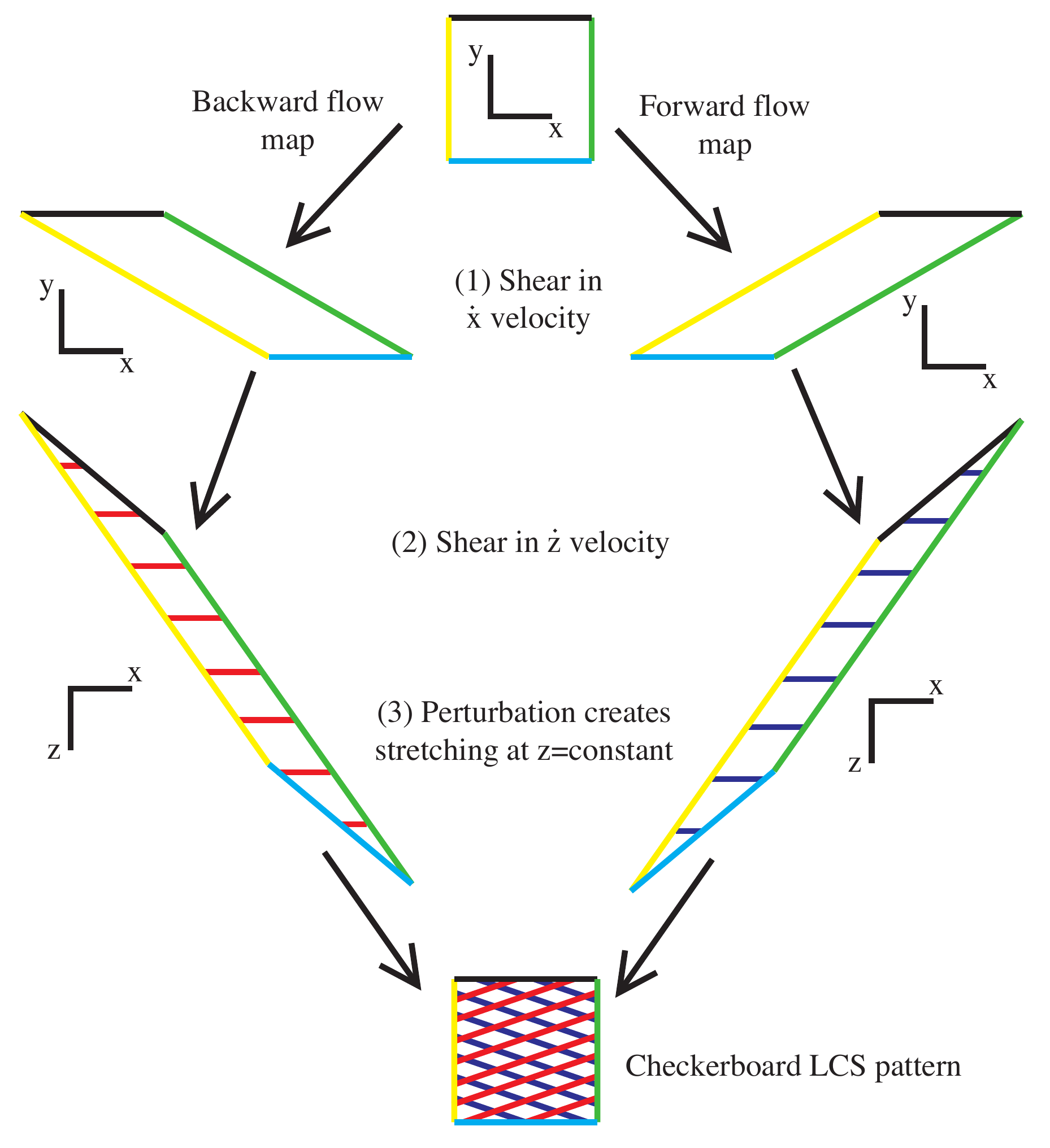}
  \caption{The stretching behavior resulting from Eq. \ref{eq:model}. An initial square at constant $z$ is stretched into a rhombus in different directions by the forward or backward flow map based on the shear in the $\dot{x}$ flow component, the $\dot{z}$ component shears the rhombus in the $z$ direction and the perturbation creates LCS at locations that have been mapped to $z(T)=$constant.\label{fig:schematic}}
\end{figure}

\begin{figure}
  \centering
  \includegraphics[width=0.5\textwidth]{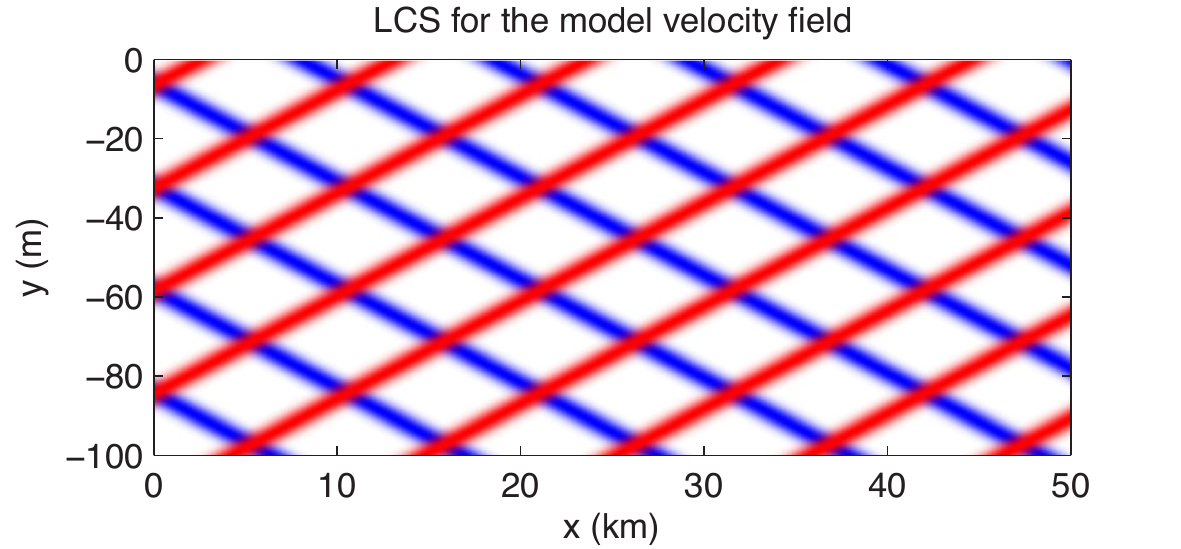}
  \caption{The LCS resulting from Eq. \ref{eq:model}. The LCS show a checkerboard pattern which is similar to the that seen in the WCR. The red, vertical curves are attracting LCS and the blue horizontal curves are repelling LCS. \label{fig:model_LCS}}
\end{figure}

To understand why this flow creates the checkerboard pattern in the LCS we consider the affect of each flow component separately. This is shown schematically in figure~\ref{fig:schematic}. Starting from an initially square domain representing the view shown in figure~\ref{fig:model_LCS}, the $\dot{x}$ component shears the square into a rhombus in opposite directions depending on whether the forward or backward flow map is considered. The $\dot{z}$ component shears the rhombus out of the $x$-$y$ plane in the $z$ direction.  Perturbations to the flow then act at positions that are mapped to $z=$constant, generating larger stretching at these locations and therefore LCS. These LCS are aligned as shown in figure~\ref{fig:schematic}, creating transversely intersecting forward and backward LCS.

The velocity field of equation~\ref{eq:model} is simple enough to be integrated analytically, but the full solution for $z(t)$ is sufficiently complex that it is most instructive to examine the flow for $\varepsilon=0$ and then consider the effect of the perturbation. If $\epsilon=0$, the flow map from $t=0$ to $t=T$ is
\begin{align}
  x(T) &= x_0+\dot{\gamma}_1y_0T \notag\\
  y(T) &= y_0 \label{eq:flowmap}\\
  z(T) &= z_0 +\dot{\gamma}_2 x_0 T + \dfrac{\dot{\gamma}_1\dot{\gamma}_2 y_0 T^2}{2}. \notag
\end{align}
Thus, material which is mapped to $z(T)$=constant and therefore acted on uniformly by the perturbations originates on the plane defined by
\begin{equation}
  z(T) = z_0 +\dot{\gamma}_2 T x_0 + \dfrac{\dot{\gamma}_1\dot{\gamma}_2  T^2}{2}y_0
\end{equation}
where $(x_0,y_0,z_0)$ are the starting coordinates. The perturbation creates compression and expansion in $z(T)$ at a wavelength of $\lambda$. Thus, we expect the resulting LCS to spaced at intervals of
\begin{equation}
  \Delta x=\dfrac{\lambda}{\dot{\gamma}_2 T},~~~\Delta y=\dfrac{2\lambda}{\dot{\gamma}_1 \dot{\gamma}_2 T^2},~~~\Delta z = \lambda,
\end{equation}
and a slope in $x$-$y$ plane of
\begin{equation}
  \dfrac{\Delta y}{\Delta x} = \dfrac{2}{\dot{\gamma}_1 T}.
\end{equation}
These results are valid for $\varepsilon\ll 1$. Larger values of $\varepsilon$ increase the spacing between the LCS, but do not change the slope.

To compare the model velocity field to the checkerboard LCS seen in the WCR we must first estimate the parameters $\dot{\gamma}_1$, $\dot{\gamma}_2$, and $\lambda$ in the WCR. $\dot{\gamma}_1$ corresponds to the shear rate of the radial velocity component with respect to the vertical direction. An examination of the azimuthally averaged WCR shows that the radial outflow component has a maximum near the surface of about $2.5\times10^{-2}$ m s$^{-1}$ and decreases to zero at a depth of about 75 m so  $\dot{\gamma}_1\approx3.33\times10^{-4}$ s$^{-1}$. $\dot{\gamma}_2$ corresponds to the shear rate of the azimuthal velocity with respect to the radius in the checkerboard region. In the same way, we find that $\dot{\gamma}_2\approx1.15\times 10^{-5}$ s$^{-1}$ in the checkerboard region. Despite the fact that the radial flow velocity in the WCR is much smaller than the azimuthal component, $\dot{\gamma}_1$ is an order of magnitude larger than $\dot{\gamma}_2$ due to the smaller length scales in the vertical direction. For this reason, the shear in the WCR radial outflow is a critical component for generating the checkerboard LCS pattern.

Finally, we determine the parameters of the model associated with the sinusoidal perturbation. The value of $\varepsilon$ does not significantly affect the resulting LCS. We use $\varepsilon=0.01$, corresponding to a 1\% perturbation of the azimuthal velocity. An examination of the WCR shape reveals that it is slightly elliptical. This is common in WCRs~\citep{Cushman:85a} and generates a perturbation wavelength of 1/2 the WCR circumference. The circumference of the checkerboard region is about $5.9\times10^{5}$ m, giving a value of $\lambda=2.95\times10^{5}$ m. Computing the LCS in this model with the same integration time that was used for the WCR ($T=\pm4$ weeks) results in the checkerboard LCS pattern shown in figure~\ref{fig:model_LCS}.

To directly compare the checkerboard LCS resulting from the model and the WCR, we examine two metrics: the spacing between the LCS and the slope of the LCS. For the WCR, the LCS have a slope of around 2-3 m km$^{-1}$, a horizontal spacing of about 8-15 km and a vertical spacing of 20-40 m depending on precisely where these measurements are made.  For the model velocity field of equation~\ref{eq:model}, the LCS have a slope of 2.5 m km$^{-1}$, a horizontal spacing of 10.6 km, and a vertical spacing of 26.6 m. These results all lie within the range of values measured in the WCR and show very good agreement between the model and the checkerboard region in the WCR.

\section{Conclusions}

A Lagrangian coherent structure (LCS) analysis of a warm-core ring (WCR) in the Gulf of Mexico has revealed some previously unobserved transport structures. In the WCR, the vortex core is surrounded by a series of ``checkerboard'' LCS in the mixed layer which form a cross-hatched pattern of transversely intersecting LCS when viewed in the $r$-$z$ plane. Fluid in this region undergoes consistent stretching behavior. As with most ocean flows, there is very little vertical transport in the WCR and a box of passive drifter particles placed in the checkerboard region is elongated in the circumferential direction, becoming wrapped around the WCR, while becoming thinner in the radial direction. Such uniform stretching behavior does not admit transport across the checkerboard LCS region, contributing to the long life of WCRs.

A detailed investigation of the Lagrangian stretching behavior in the checkerboard region gives rise to an analytical model which reproduces the checkerboard LCS. The most important parameters of the model are the vertical shear rate of the radial velocity component, the horizontal shear rate of the azimuthal velocity, and the principle perturbation wavelength of deviations from axisymmetry. These parameters were estimated directly from the WCR velocity field and used to compute the LCS generated by the model. The LCS which are produced by this analytical model show very good agreement with those seen in the WCR. Transversely intersecting LCS are produced by the model with slopes and spacings that match those seen in the WCR.

It is important to note that although the velocity field in the WCR is largely two dimensional, the length scales in the vertical direction are much smaller than the horizontal. For this reason, gradients in the vertical direction can be of the same order or larger than those in the horizontal. The model used to produce the checkerboard LCS here relies on three dimensional stretching behavior which cannot occur in a two dimensional domain. Even though the vertical motion is ignored, the vertical shear is critically important. Additionally, the small radial outflow near the surface of the WCR is absolutely critical in producing the transversely intersecting LCS pattern seen here. This radial flow is thought to occur primarily because frictional effects in the mixed layer disrupt the cyclo-geostrophic balance of the WCR and enable a pressure driven radial outflow. Below the mixed layer, there is essentially no radial flow component and therefore the character of the LCS changes to include only parallel LCS.

The checkerboard LCS structures seen in this study have been shown to be associated with the shear present in this region of the WCR. In WCRs and other well developed mesoscale ocean eddies, flow tends to be well ordered and laminar and the lack of strong updrafts and overturning flow in WCRs limits the available mechanisms for mixing and homogenization. The presence of the checkerboard LCS reveals one region in the WCR where shear greatly affects the Lagrangian dynamics of fluid motion. A study of drifters placed in this region revealed that a large box shaped domain is quickly stretched into a long, thin filament around the ring circumference. Small scale mixing caused by breaking waves, wind gusts, biological interactions, and small-scale turbulence is always present in the ocean and the stretching and shearing in the checkerboard region creates opportunities for mixing and homogenization within the WCR on small scales while minimizing fluid exchange with the surrounding ocean.

The enhanced shearing, mixing, and homogenization within the checkerboard region likely has affects on biological systems. It is known that warm-core rings influence phytoplankton blooms~\citep{Franks:86a,Biggs:92a} and fish distributions~\citep{Olson:85a} in the ocean. In fact, the advective transport in this region will impact the distribution of nutrients, pollutants, temperature, salt, oxygen, etc. within and around the WCR.

\newcommand{\AIAAJ}{AIAA J.} \newcommand{\AIAAP}{AIAA Paper}
  \newcommand{\AM}{Acta Math.} \newcommand{\ARMA}{Archive for Rational
  Mechanics and Analysis} \newcommand{\ASMEJFE}{J. Fluids Eng., Trans. ASME}
  \newcommand{\ASR}{Applied Scientific Research} \newcommand{\CF}{Computers
  Fluids} \newcommand{\ETFS}{Experimental Thermal and Fluid Science}
  \newcommand{\EF}{Experiments in Fluids} \newcommand{\FDR}{Fluid Dynamics
  Research} \newcommand{\IJHMT}{Int. J. Heat Mass Transfer}
  \newcommand{\JASA}{J. Acoust. Soc. Am.} \newcommand{\JCP}{J. Comp. Physics}
  \newcommand{\JFM}{J. Fluid Mech} \newcommand{\JMP}{J. Math. Phys.}
  \newcommand{\JSC}{J. Scientific Computing} \newcommand{\JSP}{J. Stat. Phys.}
  \newcommand{\JSV}{J. of Sound and Vibration} \newcommand{\MC}{Mathematics of
  Computation} \newcommand{\MWR}{Monthly Weather Review}
  \newcommand{\PAS}{Prog. in Aerospace. Sci.} \newcommand{\PCPS}{Proc. Camb.
  Phil. Soc.} \newcommand{\PD}{Physica D} \newcommand{\PRA}{Physical Rev. A}
  \newcommand{\PRE}{Physical Rev. E} \newcommand{\PRL}{Phys. Rev. Lett.}
  \newcommand{\PF}{Phys. Fluids} \newcommand{\PFA}{Phys. Fluids A.}
  \newcommand{\PL}{Phys. Lett.} \newcommand{\PRSLA}{Proc. R. Soc. Lond. A}
  \newcommand{\SIAMJMA}{SIAM J. Math. Anal.} \newcommand{\SIAMJNA}{SIAM J.
  Numer. Anal.} \newcommand{\SIAMJSC}{SIAM J. Sci. Comput.}
  \newcommand{\SIAMJSSC}{SIAM J. Sci. Stat. Comput.}
  \newcommand{\TCFD}{Theoret. Comput. Fluid Dynamics} \newcommand{\ZAMM}{ZAMM}
  \newcommand{\ZAMP}{ZAMP} \newcommand{\ICASER}{ICASE Rep. No.}
  \newcommand{\NASACR}{NASA CR} \newcommand{\NASATM}{NASA TM}
  \newcommand{\NASATP}{NASA TP} \newcommand{\ARFM}{Ann. Rev. Fluid Mech.}
  \newcommand{\WWW}{from {\tt www}.} \newcommand{\CTR}{Center for Turbulence
  Research, Annual Research Briefs} \newcommand{\vonKarman}{von Karman
  Institute for Fluid Dynamics Lecture Series}

\end{document}